\pdfoutput=1

\documentclass[10pt]{article}

\usepackage{listings}    
\usepackage{graphicx}    
\usepackage{subcaption}  
\usepackage{algorithm}   
\usepackage{algorithmic} 
\usepackage{booktabs}    
\usepackage{hyperref}    
\usepackage{xurl}        
\usepackage[htt]{hyphenat} 
\usepackage{microtype} 
\DisableLigatures{}    
\usepackage{adjustbox} 
\usepackage{colortbl}  

\usepackage{caption}
\usepackage{setspace}
\usepackage{multirow}
\usepackage{enumerate}
\usepackage{pdflscape}
\usepackage{pgfplots} 

\usepackage[table]{xcolor}
\definecolor{codegreen}{rgb}{0.25,0.5,0.35}
\definecolor{codegray}{rgb}{0.5,0.5,0.5}
\definecolor{codepurple}{rgb}{0.6,0,0}
\definecolor{backcolour}{rgb}{0.95,0.95,0.92}
\definecolor{colorstring}{rgb}{0.5,0,0.35}
\definecolor{rltred}{rgb}{0.5,0,0}
\definecolor{rltgreen}{rgb}{0,0.5,0}
\definecolor{rltblue}{rgb}{0,0,0.5}
\definecolor{DarkGreen}{rgb}{0.00,0.60,0.00}
\definecolor{ScarletRed}{rgb}{0.80,0.00,0.00}
\definecolor{blizzardblue}{rgb}{0.67, 0.9, 0.93}
\definecolor{green-yellow}{rgb}{0.68, 1.0, 0.18}
\definecolor{dkgreen}{rgb}{0,0.6,0}
\definecolor{gray}{rgb}{0.5,0.5,0.5}
\definecolor{mauve}{rgb}{0.58,0,0.82}
\definecolor{lightgrey}{rgb}{0.90,0.90,0.90}
\definecolor{grey}{gray}{0.75}
\definecolor{light-gray}{gray}{0.80}

\lstdefinestyle{mystyle}{
    escapechar=©, 
	backgroundcolor=\color{backcolour},
    basicstyle=\scriptsize\ttfamily,
   	identifierstyle=\footnotesize\ttfamily,
	commentstyle=\color{codegreen},
	keywordstyle=\color{colorstring}\bfseries,
	numberstyle=\ttfamily\color{codegray},
	stringstyle=\ttfamily\color{DarkGreen},
	breakatwhitespace=false,
	breaklines=true,
	captionpos=b,
	keepspaces=true,
	numbers=left, 
	numbersep=2pt,
	showspaces=false,
	showstringspaces=false,
	showtabs=false,
	tabsize=2
}
\usepackage{xspace}
\newcommand{\evo}{{\sc EvoMaster}\xspace}

\usepackage{boxedminipage}
{\smallskip
	\noindent
	\let\emph=\textbf
	\begin{boxedminipage}{\columnwidth}\begin{center}\em}%
		{\end{center}\end{boxedminipage}%
}

\usepackage{ifthen}
\newboolean{showcomments}
\setboolean{showcomments}{true} 

\ifthenelse{\boolean{showcomments}}{
	\newcommand{\nbc}[3]{
		{\colorbox{#3}{\bfseries\sffamily\scriptsize\textcolor{white}{#1}}}
		{\textcolor{#3}{\sf\small$\langle$\textit{#2}$\rangle$}}}
	
}{
	\newcommand{\nbc}[3]{}

}

\usepackage{authblk}                
\usepackage[square,numbers]{natbib} 
\usepackage{amsmath}                
\usepackage{amssymb}                

\usepackage{geometry}
\title{
Using OAI Overlay to Enhance REST API Fuzzing
}

\author[1,2,8]{Omur Sahin}
\author[3]{Man Zhang}
\author[4]{Alexander Poth}
\author[4]{Olsi Rrjolli}
\author[5]{Andreas Faes}
\author[6]{Piyun Teng}
\author[6]{Kaiming Xue} 
\author[9]{Wenjuan Ma} 
\author[2,7]{Andrea Arcuri}

\affil[1]{Erciyes University, Türkiye}
\affil[2]{Kristiania University College, Norway}
\affil[3]{Beihang University, China}
\affil[4]{Volkswagen AG, Germany}
\affil[5]{Independent, Belgium}
\affil[6]{Meituan, China}
\affil[7]{Oslo Metropolitan University, Norway}
\affil[8]{UniMetrik Teknoloji Ltd. Şti., Türkiye}
\affil[9]{Independent, China}

\date{}

\begin{document}

\maketitle

\begin{abstract}
REST APIs are widely used in industry.
Therefore, a lot of research has been focused on how to automatically generate test cases for REST APIs, with few different open-source fuzzers existing in the literature.
For a thorough testing, especially in black-box scenarios, just relying on the information provided in the OpenAPI schemas is not enough.
Testers typically need to provide extra input data to help steering the fuzzers in the right direction.
Dedicated formats specific to each different fuzzer would work, but they would create a vendor lock-in, as well as increasing cognitive load.
The OpenAPI Initiative (OAI) standard Overlay might be a solution to this problem.
Such standard enables to define transformations on the OpenAPI schemas, where testers can provide input data in Overlay files where such data is provided as ``examples'' entries.
In this paper, we have extended the state-of-the-art fuzzer \evo to support Overlay files natively.
Experiments are carried out in industry on five APIs from five enterprises from around the world (e.g., Belgium, China, Germany and Türkiye), including two Fortune500 enterprises as well as a 3-man startup.
Our industrial results show that Overlay is a viable solution to better enable black-box fuzzing of REST APIs in industry.

\end{abstract}

{\bf Keywords}: REST, API, fuzzing, black-box, testing

\section{Introduction}

In black-box testing of REST APIs, the main source of information for a fuzzer would be the schema
(e.g., expressed in OpenAPI\footnote{\url{https://spec.openapis.org/oas/latest.html}} format)
and the feedback from responses of its HTTP calls towards the tested API (e.g., status codes and body payloads).

With only this type of information available, it might  not be possible for a fuzzer to achieve high testing effectiveness.
This would be due to difficulty of creating the right test data and determining the right dependencies between operations.
There are several existing fuzzers in the literature~\cite{golmohammadi2022testing}, with significant research progress, but still none achieve high enough test quality to fully replace a human tester~\cite{Kim2022Rest,zhang2023open,sahin_2025_wfc}.

Still, existing fuzzers can be of practical value for practitioners in industry~\cite{icst2025vw,zhang2025fuzzing,corradini2026automated}.
Even if they cannot fully automate the generation of all needed testing scenarios, they can speed up the work of the test engineers, taking care of at least the easy, tedious parts, and possibly as well finding new, unexpected results.
Once a fuzzer has completed its job, still the test engineer has to do the manual work to test all the other remaining uncovered scenarios.

If the engineer/tester still has to do the work of writing the missing tests manually, the question is if it would payoff to use some of that time to provide ``hints'' to the fuzzer.
These could be for example data for different query parameters, as well as specific object payloads (e.g., in JSON format) with complex constraint dependencies among parameters that are not specified in the schema.
Or provide valid ids (e.g., in UUID format) for existing data in the test environment's databases.

Currently, most REST API fuzzers already can support this feature, as they can use ``example'' and ``examples'' entries in the OpenAPI schema when generating input data.
The problem, though, is that this would require to modify the schema with test data.
This is not viable in the long run.
As soon as there is a new update in the schema, all the examples would need to be manually re-added to the new schema.
This is particular the case when the tester of the API is not the designer of the schema itself.
Also, in large schemas with hundreds of HTTP endpoints, adding example entries for each functionality could be unmanageable.

The alternative to use a separated, custom format to provide input data hints to the fuzzer is not a particularly good option either.
It would create a ``vendor-lockin'' for each specific fuzzer.
Also, IDE support for such format (e.g., autocomplete and validation) could be limited, besides the time needed to invest in learning a new custom format.
Furthermore, LLM-support for learning and creating such files would likely be limited as well, due to lack of instance examples during its training.

To address this software engineering problem of practical value in industry,
a possible solution is to use Overlay.\footnote{\url{https://spec.openapis.org/overlay/latest.html}}
At the time of writing, this is a recent (2024) specification standard developed by the OpenAPI Initiative (OAI).
It provides a way to define transformation rules that can be applied to an existing OpenAPI schema.

In this paper, we propose and evaluate how such Overlay transformations can be used to add ``examples'' entries in the OpenAPI schema of the tested API.
In the rest of the paper, the term ``we'' refers to the development team of \evo~\cite{arcuri2025tool}.
Industrial partners and collaborators that joined in this work only ran and analyzed the experiments on their industrial APIs.
They were not involved in any coding nor technique design.

By relying on OpenAPI's handling of ``named'' examples introduced in version 3.1.0, it is also possible to specify values for combinations of parameters by giving the same example name, as long as the fuzzer can recognize those cases.
If this approach is viable, with no major drawback, it would solve the addressed problem, as Overlay is a standard, used also outside of software testing needs.
The transformations could be applied on an OpenAPI schema before giving it as input to the fuzzer (e.g., using any existing Overlay merge tool).
Alternatively, if a fuzzer has ``native'' support for Overlay, then it would just be a matter of giving the Overlay file(s) as input besides the OpenAPI schema.
This would simplify its use, as there would be no need to install and run a second tool besides the fuzzer.

For our empirical study, we added native support for Overlay to the state-of-the-art fuzzer \evo~\cite{arcuri2025tool}.
We implemented a JVM-based library (written in Java) called \texttt{overlay-jvm}, which we use internally in \evo to apply Overlay transformations.
This library is released open-source on GitHub\footnote{\label{link:overlay-jvm}\url{https://github.com/WebFuzzing/overlay-jvm}} and Maven.

For this feasibility study,
five
different enterprises worldwide were involved, varying in size and business category.
These include
Fortune500 enterprises such as the car maker Volkswagen AG (Germany) and e-commerce platform Meituan (China),
the 3-man startup UniMetrik (Türkiye),
a logistics service provider which we are currently not authorized to name (Belgium),
and a fifth enterprise for which we cannot provide any info about for legal reasons.

Test engineers in each of these enterprises were tasked to choose one of their APIs for these experiments, and write an Overlay transformation file to add meaningful test data examples to the OpenAPI schema.
They were then asked to run \evo with and without Overlay support, and compare the results.

From the point of view of the fuzzer, this is not different than making a copy of the OpenAPI schema and adding ``examples'' entries manually directly in this copy before giving it as input to the fuzzer, as for example done in~\cite{icst2025vw}.
Adding examples has already shown to be useful to drive fuzzers to get better results, where the obtained benefits outweigh the manual costs for the testers~\cite{icst2025vw}.
In this paper, we rather aim at studying whether the use of Overlay for this task is viable and potentially a better, more scalable and maintainable approach for practitioners in industry.

The rest of the paper is organized as follows.
Section~\ref{sec:relatedwork} discusses related work.
Section~\ref{sec:overlay} shows an example of use for Overlay.
How the support for Overlay was integrated in the fuzzer \evo is discussed in Section~\ref{sec:integration}.
Our empirical study in industry is presented in Section~\ref{sec:study}.
A discussion on threats to validity follows in Section~\ref{sec:threats}.
Finally, Section~\ref{sec:conclusions} concludes the paper.


\section{Related Work}
\label{sec:relatedwork}

Several REST API fuzzers have been introduced in the literature~\cite{golmohammadi2022testing}, including for example
APIRL~\cite{foley2025apirl},
ARAT-RL~\cite{kim2023adaptive},
AutoRestTest~\cite{kim2025autoresttest},
EmRest~\cite{xu2025effective},
\evo~\cite{arcuri2025tool},
Morest~\cite{liu2022icse},
QuickREST~\cite{karlsson2020QuickREST},
RestCT~\cite{wu2022icse},
RESTest~\cite{martinLopez2021Restest},
Restler~\cite{restlerICSE2019},
RestTestGen~\cite{viglianisi2020resttestgen},
Schemathesis~\cite{hatfield2022deriving},
and
WuppieFuzz~\cite{rooijakkers2025wuppiefuzz}.

Many of these fuzzers are  academic prototypes, that have been discontinued and are no longer maintained.
Based on their popularity on open-source host services such as GitHub (e.g., based on number of ``stars'' and download statistics),
notable exceptions are though Schemathesis~\cite{hatfield2022deriving}, Restler~\cite{restlerICSE2019} and our own work \evo~\cite{arcuri2025tool}.

Most of the existing studies in the scientific literature on REST API fuzzing seem to have been mainly based on lab experiments, with no involvement with industrial practitioners.
There are only a few exceptions.
These include the work done with QuickREST at ABB~\cite{karlsson2020QuickREST},
Morest at Huawei~\cite{liu2022moresthuawei},
and RestTestGen at the space solutions company SES~\cite{corradini2026automated}.
In our previous work,
the very first experiments with \evo were done with a company we are not authorized to name~\cite{arcuri2017restful},
followed up by collaborations with Meituan~\cite{zhang2023rpc,zhang2024seeding,zhang2025fuzzing},
as well as with Volkswagen~\cite{poth2025technology,icst2025vw,garrett2025generating,arcuri2026fuzzing}.

To the best of our knowledge, no existing study in the literature has proposed and evaluated the use of Overlay to enhance the use of black-box REST API fuzzing in industry.
By involving practitioners and industrial APIs from
five
different enterprises worldwide, the work in this paper represents the most variegated study in industry to date.

\section{Overlay Example}
\label{sec:overlay}

\begin{figure}[!t]
\begin{lstlisting}
openapi: 3.1.0
info:
  title: Simple API
  version: 1.0.0
servers:
  - url: https://api.example.com/v1
paths:
  /api:
    post:
      parameters:
        - name: x
          in: query
          schema:
            type: string
        - name: y
          in: query
          schema:
            type: string
      requestBody:
        required: true
        content:
          application/json:
            schema:
              type: object
              properties:
                id:
                  type: string
                name:
                  type: string
      responses:
        '200':
          description: Successful response
\end{lstlisting}
\caption{\label{fig:schema}
Artificial example of an OpenAPI schema defining one \texttt{POST} endpoint.
}
\end{figure}

Figure~\ref{fig:schema} shows a simple, artificial example of an OpenAPI schema written in version $3.1.0$.
It contains only one single endpoint with a \texttt{POST}.
This operation has two query parameters \texttt{x} and \texttt{y}, as well as a body payload being an object with parameters \texttt{id} and \texttt{name}.

\begin{figure}[!t]
\begin{lstlisting}
overlay: 1.1.0
info:
  title: Add Examples to API
  version: 1.0.0
actions:
  - target: '$.paths["/api"].post.parameters[?(@.name=="y")]'
    description: Add foo example to query parameter y
    update:
      examples:
        foo:
          value: Foo
  - target: '$.paths["/api"].post.parameters[?(@.name=="x")]'
    description: Add bar example to query parameter x
    update:
      examples:
        bar:
          value: Bar
  - target: '$.paths["/api"].post.parameters[?(@.name=="y")]'
    description: Add bar example to query parameter y
    update:
      examples:
        bar:
          value: Bar
  - target: '$.paths["/api"].post.requestBody.content["application/json"]'
    update:
      examples:
        bar:
          value:
            id: "bar-id-123"
            name: "Bar Name Example"
\end{lstlisting}
\caption{\label{fig:overlay}
Overlay file definition used to add four ``examples'' entries.
}
\end{figure}

Assume a test engineer wants to make sure some specific values are used for some of these parameters, or combinations of them.
These can be added via ``examples'' entries.
Figure~\ref{fig:overlay} shows an Overlay file definition in which four transformation \texttt{actions} are defined to add \texttt{examples} entries.
Each ``action'' in the \texttt{actions} list requires to define the \texttt{target} node(s) in the schema that need to be modified, using an RFC9535 JSONPath query expression.
Then, three different types of mutually-exclusive operations are available: \texttt{update}, \texttt{copy} and \texttt{remove}.

\begin{figure}[!t]
\begin{lstlisting}
openapi: 3.1.0
info:
  title: Simple API
  version: 1.0.0
servers:
  - url: https://api.example.com/v1
paths:
  /api:
    post:
      parameters:
        - name: x
          in: query
          schema:
            type: string
          examples:
            bar:
              value: Bar
        - name: y
          in: query
          schema:
            type: string
          examples:
            bar:
              value: Bar
            foo:
              value: Foo
      requestBody:
        required: true
        content:
          application/json:
            schema:
              type: object
              properties:
                id:
                  type: string
                name:
                  type: string
            examples:
              bar:
                value:
                  id: "bar-id-123"
                  name: "Bar Name Example"
      responses:
        '200':
          description: Successful response
\end{lstlisting}
\caption{\label{fig:transformed}
Schema result of applying Overlay transformation in Figure~\ref{fig:overlay} on the schema defined in Figure~\ref{fig:schema}.
}
\end{figure}

Figure~\ref{fig:transformed} shows the results of applying the Overlay transformation in Figure~\ref{fig:overlay} on the schema defined in Figure~\ref{fig:schema}.
Two different ``named'' examples are added: \texttt{foo} and \texttt{bar}.
The example named \texttt{foo} defines a string instance for the query parameter \texttt{y}.
On the other hand, the name \texttt{bar} is used for adding  example entries to all query parameters and body payload.

\section{Fuzzer Integration}
\label{sec:integration}

We have extended the state-of-the-art~\cite{Kim2022Rest,sartaj2025rest,sahin_2025_wfc,arcuri2026sbft} fuzzer \evo to support Overlay transformations.
This is done in a Java library called \texttt{overlay-jvm}, which we have open-sourced on GitHub.\footref{link:overlay-jvm}
As this library can be used outside of fuzzing and software testing needs, we have developed it and released independently from \evo.
At the time of writing, there was no existing library for OAI Overlay support for the JVM that was updated and still maintained.
With this support, testers and engineers in industry can provide test data for the fuzzers in separated Overlay transformation files, using a standardized format.

A state-of-the-art fuzzer such as \evo is able to handle and exploit ``examples'' entries, as most REST API fuzzers do.
Each time a fuzzer needs to sample a new test case to evaluate, each entry can be either sampled at random (but still keeping with the constraints defined in the schema, like pattern regex and min/max ranges), or from a pre-defined dictionary of values.
If there are ``examples'' in the schema, those can be simply used as a further pool of values to sample from.

But what if a tester wants to make sure a fuzzer is evaluating a specific combination of values for different parameters?
For example, considering the schema in Figure~\ref{fig:schema}, and having \texttt{x=Bar}, \texttt{y=Bar} while the payload is the JSON object \texttt{\{id: "bar-id-123", name: "Bar Name Example"\}}.
This can be achieved by using the same \emph{name} for the example (i.e., \texttt{bar}), and making sure the fuzzer can recognize these cases.
The use of named examples were introduced in version 3.0.0 of OpenAPI, and refined in version 3.1.0 (by allowing to have them also under Schema Object entries).

In the case of \evo, we have extended it with the following algorithm.
If during sampling of a test case any example value is chosen, and such example value is named, and for this name there is more than one parameter/body/field having an example value with the same name, then with a given probability $P$ (e.g., $P=0.5$) all those other entries are automatically modified using their example value with that same name.
Note that, given the same name, different properties can have different values for the named examples, like  \texttt{x=Bar} and \texttt{\{id: "bar-id-123", name: "Bar Name Example"\}} for the name \texttt{bar}.

During the fuzzing process, hundreds of thousands of HTTP calls could be evaluated, especially if the fuzzer session is left running for hours.
Generating output test suites with hundreds of thousands of HTTP calls would not be viable.
As such, fuzzers would generate as output only a minimized test suite, based on different testing criteria.
These could be for example based on black-box REST API schema coverage~\cite{martin2019test} and detected distinct faults.
Still, this does not guarantee that test cases using the chosen example values would end up in the final output test suite.

To address this problem, \evo already creates testing targets for each example combined with each distinct returned status code~\cite{icst2025vw,arcuri2026fuzzing}.
Given the example value \texttt{x=Bar}, \evo  not only makes sure that it keeps in the final test suite at least one test case that use such example, but also for each distinct HTTP status code \evo obtained when using it (e.g., 200 and 400).
In this paper, we extended \evo to apply the same approach for the combinations of parameter values having the same example name.
We create a new testing target for each named example combination based for each different  returned HTTP status code.
This enables \evo to retain this type of test in the final output test suite.
Note that \evo has an advanced archive system in which evolved tests during the fuzzing are optimized.
For example, shorter tests that cover more testing targets are prioritized, and tests are continuously replaced in the archive when better tests are generated.
Still, once a testing target has been covered with a new test case, \evo guarantees that at least one test case will cover such target in the final output test suite.

\begin{figure}[!t]
\includegraphics[width=1.0\textwidth]{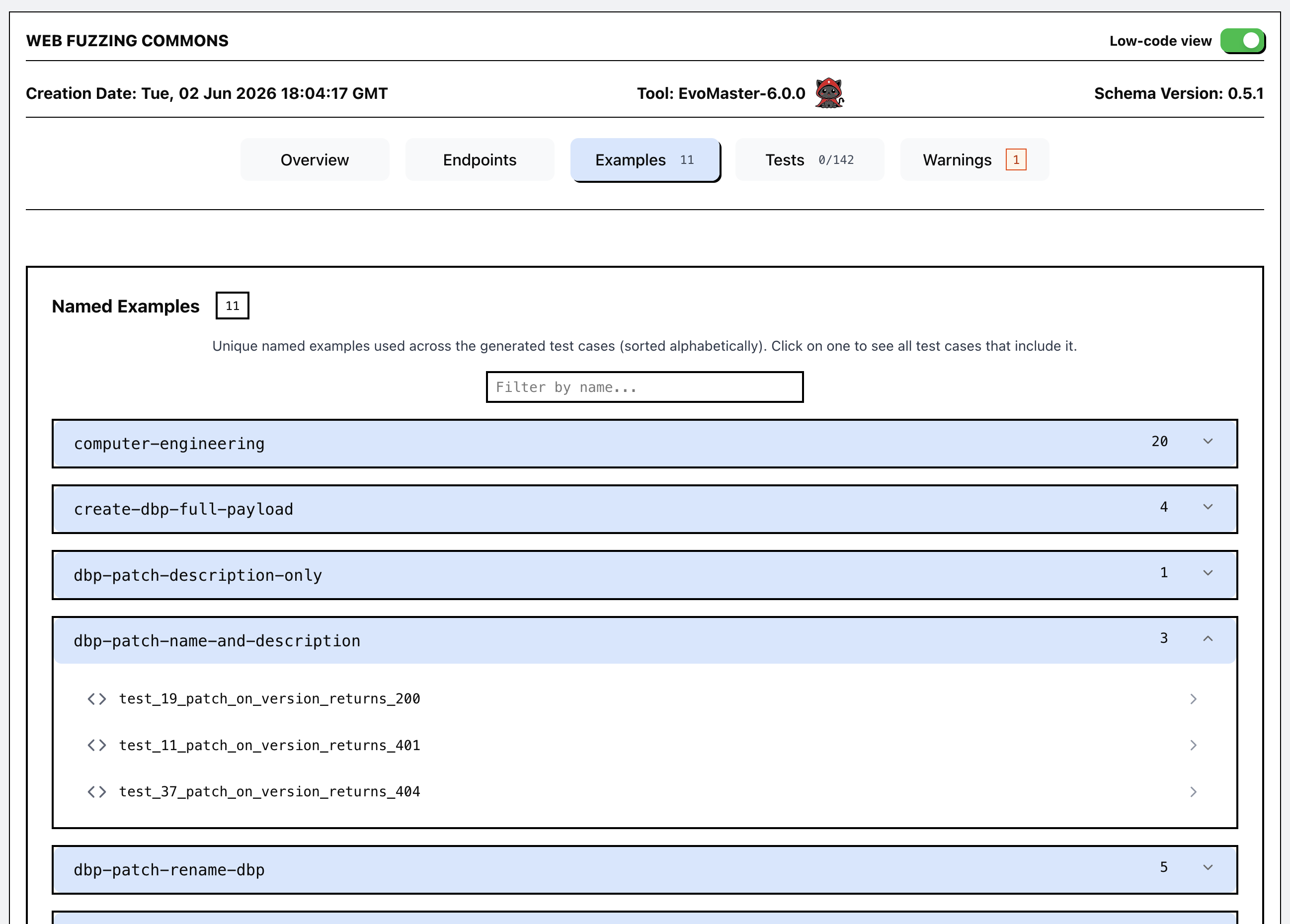}
\caption{
\label{fig:wfc}
Screenshot of WFC Web Report opened on the ``Examples'' view.
}
\end{figure}

Regardless of what code and HTTP coverage has been achieved by the generated test suites, test engineers might still need to be able to easily ``trace'' how the provided examples are used.
This is particularly important in industry where generated test suites might need to be reviewed and discussed with different stakeholders and product owners.
Besides generating test suites in different programming languages (i.e., Java, Kotlin, Python and JavaScript), \evo also creates as output an interactive web test report, using WFC Web Report~\cite{sahin_2025_wfc}.
In this paper, we have extended such report by creating a new view for easily trace where each named example is used in which generated test.
Figure~\ref{fig:wfc} shows a screenshot of a web report opened on that view.

\section{Empirical Study}
\label{sec:study}

To evaluate the use of Overlay to address the problem of providing input data to REST API fuzzers, we carried out an empirical study in industry.
As previously mentioned,
five
different enterprises have been involved.
These are
Fortune500 enterprises such as the car maker Volkswagen AG\footnote{\url{https://www.volkswagen.com}} (Germany)
and e-commerce platform Meituan\footnote{\url{https://www.meituan.com}} (China),
the 3-man startup UniMetrik\footnote{\url{https://www.unimetrik.com}} (Türkiye),
a logistics service provider (Belgium),
and a fifth enterprise that, for legal reasons, requested us to not disclose any information (e.g., name and business type).

The enterprises Volkswagen AG and Meituan have been users of \evo for some years, and have been involved before in studies involving \evo~\cite{zhang2023rpc,zhang2024seeding,zhang2025fuzzing,poth2025technology,icst2025vw,garrett2025generating,arcuri2026fuzzing}.
On the other hand, the other enterprises are new, first-time users of \evo.

Each enterprise received the same set of instructions to conduct these experiments on their premises.
These instructions can be summarized as follows:
\begin{itemize}
\item Choose one of their industrial APIs.
\item Write an Overlay file to add meaningful test data for that chosen API.
\item Setup any needed authentication information (e.g., in WFC format~\cite{sahin_2025_wfc}).
\item Run \evo (latest version 6.0.0) with and without Overlay support on such API, for 10 minutes.
\item Collect the results of the experiments, e.g., 2xx HTTP status coverage per endpoint and the number of detected faults.
\item Provide, in writing, any positive or negative comment about the experience of writing Overlay files.
\end{itemize}

As a state-of-the-art tool, tool usability is of paramount importance, especially if aiming at a wider adoption in the software engineering community.
Once the Overlay files are implemented, running these experiments is as simple as running:

\begin{lstlisting}[basicstyle=\normalsize\ttfamily,]
> pip install evomaster
>
> evomaster --schema $SCHEMA --base $BASE --maxTime 10m --overlay $OVERLAY
\end{lstlisting}

where \texttt{\$SCHEMA} is the location of where the OpenAPI schema can be found;
\texttt{\$BASE} is an optional parameter to specify where the API is up and running (e.g., \texttt{http://localhost:8080}) if the information in the schema is pointing to the production server and not the used testing environment;
\texttt{\$OVERLAY} specifies where the Overlay file is located.
If the API needs authentication, this can be either provided with \texttt{--header0}, or put in an \texttt{em.yaml} WFC configuration file (whose default location is the current working directory, which can be changed with \texttt{--configPath} if needed).
Such ease of use is what makes possible to have empirical studies in industry using \evo in several different enterprises.

\begin{table}[!t]
    \centering
    \caption{ \label{table:results}
    Results for each API, where we list the number of their endpoints (\#Endp) and the number of actions/examples in the written Overlay files (\#Overlay).
    We report 2xx coverage and fault detection for when the base version of \evo is used (B-2xx and B-Faults),
    compared to when Overlay support is employed (O-2xx and O-Faults).
    }
\begin{tabular}{ l rr | rr | rr   }\\
\toprule
API & \#Endp & \#Overlay & B-2xx & B-Faults & O-2xx & O-Faults \\
\midrule
A0 &  3 & 18 & 1 & 3 &  3 & 3 \\
A1 & 16 & 19 & 7 & 0 & 11 & 3 \\
A2 & 23 & 20 & 6 & 0 & 16 & 0 \\
A3 &  6 &  8 & 1 & 8 &  2 & 9 \\
A4 &  2 &  2 & 2 & 4 &  2 & 4 \\
\bottomrule
\end{tabular}
\end{table}

With ``2xx HTTP status coverage'' we mean the number of endpoints for which at least one test case was generated that returned a status code in the range $200-299$.
However, an exception is $A2$, which does not follow HTTP semantics.
All of its responses return a 200 status code, where the success/failure  is specified in an internal \texttt{code} field returned in each JSON body in these responses.
For these cases, for this API we manually looked at the generated test cases and counted then number of endpoints in which at least one \texttt{code} representing success was returned.

Regarding fault detection~\cite{marculescu2022faults}, \evo uses several different automated oracles to detect faults.
These include checking for responses returning a 500 HTTP status code (server error),
mismatches between responses and what declared in the OpenAPI schemas,
besides several different types of security properties and injection attacks~\cite{arcuri2025fuzzing,sahin2026enhancing}.

Table~\ref{table:results} shows details about these
five
APIs and the results achieved.
For confidentiality, we decided to not specify which API belongs to whom.
We rather use anonymized labels $A0$ to $A4$ to refer to them.

Based on the results shown in Table~\ref{table:results}, in all cases it was possible to write Overlay files that lead the fuzzer to achieve better results.
How much improvement can be obtained, though, strongly depends on the quality and quantity of the provided examples.
Furthermore, there are other sources of complexity in fuzzing that would not be addressed with just providing example entries, like dependencies between operations and complex workflows.
In theory, this could still be addressed with Overlay, by adding \texttt{link} entries to the schema, if needed.
But this possibility is not something studied in this work.

When it comes to industrial feedback, it was overall positive, with statements such as:
``\emph{overlays help to separate concerns and/or realize ownership and can be useful to structure test data sets}''
and ``\emph{The overlay is quite easy to use once you get used to it}''.
Getting used to a new standard (i.e., Overlay) might take some initial learning.
It is then not surprising that in some cases practitioners used LLMs to draft first versions of these Overlay files:
``\emph{Using a large language model to generate overlay files in the required format is also relatively convenient}''.

In the case of $A3$, where improvements were  small, the feedback was: ``\emph{ I see a better result, but with the given overlays I used there is no significant gain in the 2xx test cases performed (but that is probably/mainly because I need to add more examples)}''.
In this case, though, the issue was in \evo itself, where there was a fault related to the handling of body payloads in \texttt{PATCH} requests, leading to a third of the API endpoints only returning a 415 (Unsupported Media Type).
This is unrelated to the use of Overlay and of example entries.

The API $A4$ was the smallest among the considered APIs.
On this API, already without any Overlay it was possible to successfully cover all endpoints.
Again, this is unrelated to the use of Overlay and of example entries.

Considering these cases of $A3$ and $A4$,
it is important to stress out that the use of Overlay to specify example entries is only meant to ``help'' the fuzzer to generate better results (e.g., as done for $A0$, $A1$, and $A2$, and partially $A3$).
Still, such results depend on the quality and effectiveness of the fuzzer itself.
The use of Overlay does not fix issues in the fuzzer (e.g., $A3$), and might be unnecessary if the fuzzer is already able to successfully test the target API (e.g., $A4$).
Furthermore, a feedback comment was: ``\emph{manually modifying each input parameter value to match business logic is cumbersome}''.
This should not be needed ``for each input parameter''.
Fuzzing research is still required to try to automate as much as the testing process as possible.
Until full automation is achieved (if that will even be possible), Overlay provides a convenient way to supply domain knowledge to the used fuzzers in a standardized way.

In the industrial feedback, there were also some general concerns like: ``\emph{It will be difficult to implement for very large endpoint sets}''.
For this latter point, there is no need to have only a single Overlay file.
Transformations can be written in multiple files, organized around API functionalities to test.
When \evo is run with parameter \texttt{--overlay}, this can either specify a single Overlay file to apply, or a folder that will be recursively scanned for Overlay files.
All found Overlay files will be applied.

\section{Threats To Validity}
\label{sec:threats}

The fuzzer \evo and the library \texttt{overlay-jvm} are open-source.
Anyone can review them.
Albeit they are intensively tested, we cannot guarantee they have no software issue that impacted the validity of the study presented in this paper.

We evaluated the feasibility of applying Overlay to address the issue of providing input data to REST API fuzzers.
As this was based on human activity, it is influenced by the experience and expertise of the involved subjects.

To avoid potentially limited impact, our experiments were not carried out in the lab, but rather in industry, with industrial practitioners, using industrial APIs.
Albeit the involvement of
five
different enterprises from different parts of the world makes this study one of the most variegated in the literature (at least when it comes to fuzz REST APIs), there are still threats to external validity.
There is no guarantee that our results would generalize to other enterprises and APIs tested in other business sectors.
Furthermore, there is a risk that, even within a large enterprise, the results cannot be generalized.
For example, the evaluated sample API from Volkswagen AG does not guarantee that it is representative of all Group IT services and systems at Volkswagen AG.

\section{Conclusions}
\label{sec:conclusions}

In this paper, we have evaluated the use of OAI Overlay to specify input test data to enhance black-box REST API fuzzing.
The open-source fuzzer \evo has been extended to support Overlay files natively.

Our experiments carried out on
five
APIs from
five
different enterprise worldwide show that the use of Overlay is a practical approach that can enhance black-box REST API fuzzing.
To the best of our knowledge, this is the first work in the literature that proposes and evaluates the use of Overlay to address this industry-relevant problem.

Practitioners in industry can already use Overlay to provide input test data, regardless of the choice of which fuzzers to use, as long as such fuzzer can make use of ``examples'' entries in the schema.
However, a native support of Overlay directly in the fuzzer (as done in \evo) can improve usability, as there would be one less tool to install, configure and use.
Advanced handling of ``examples'' entries can also further improve the quality of the output generated tests.

Future work will aim at improving and simplifying other ways in which  domain expert knowledge can be provided as input to the fuzzers to boost their performance on the tested APIs.

\section*{Data Availability}

All the techniques presented in this paper are implemented as part of \evo.
The fuzzer \evo is open-source on GitHub,\footnote{\url{https://github.com/WebFuzzing/EvoMaster}}
where each new release is automatically published on Zenodo for long-term storage (e.g.,~\cite{zenodo502evomaster}).
The Overlay transformation support is implemented in a library called \texttt{overlay-jvm}, open-source on GitHub\footref{link:overlay-jvm} as well.

Due to confidentiality and non-disclosure agreements, we cannot provide a replication package including the industrial APIs and Overlay files used in our study.
This is a common case for industrial studies.

\section*{Acknowledgments}
Andrea Arcuri is funded by the European Research Council (ERC) under the European Union’s Horizon 2020 research and innovation programme (EAST project, grant agreement No. 864972).
Omur Sahin is supported by the TÜBİTAK 2219 International Postdoctoral Research Fellowship Program (Project ID: 1059B192300060).
Man Zhang is supported by the National Science Foundation of China (grant agreement No. 62502022).


\bibliographystyle{ACM-Reference-Format} 

%




\end{document}